\documentclass[ol,twocolumn,showpacs,preprintnumbers,amsmath,amssymb,superscriptaddress,groupedaddress,footinbib]{revtex4-1}

\pdfoutput=1

\usepackage[export]{adjustbox}
\usepackage{amsfonts,amsmath,times,mathrsfs,graphicx,dcolumn,bm,wasysym,xcolor,epsfig,amssymb,lipsum,color,epstopdf,bbold}
\usepackage[makeroom]{cancel}
\usepackage[english]{babel}
\usepackage[T1]{fontenc}
\usepackage[colorlinks,bookmarks=true,citecolor=red,linkcolor=blue,urlcolor=blue]{hyperref}
\usepackage{hyperref}
\usepackage[tight, FIGTOPCAP, hang, raggedright, nooneline]{subfigure}

 \definecolor{Green}{RGB}{80,182,0}
\begin{document}
\title{Quantitative phase imaging based on Fresnel diffraction from a phase plate}
\author{Samira Ebrahimi}
\affiliation{Department of Physics, Shahid Beheshti University, Evin, Tehran 19839-69411, Iran}
\author{Masoomeh Dashtdar}
\email{m-dashtdar@sbu.ac.ir}
\affiliation{Department of Physics, Shahid Beheshti University, Evin, Tehran 19839-69411, Iran}

\maketitle

\textbf{The structural complexity and instability of many interference phase microscopy methods are the major obstacles toward high-precision phase measurement. In this vein, improving more efficient configurations as well as proposing new methods are the subjects of growing interest. Here we introduce Fresnel diffraction from a phase step to the realm of quantitative phase imaging. By employing Fresnel diffraction of a divergent (or convergent) beam of light from a plane-parallel phase plate, we provide a viable, simple and compact platform for three-dimensional imaging of micron-sized specimens. The recorded diffraction pattern of the outgoing light from an imaging system in the vicinity of the plate edge can be served as a hologram, which would be analyzed via Fourier transform method to measure the sample phase information. The period of diffraction fringes is adjustable simply by rotating the plate without the reduction of both field of view and fringe contrast. The high stability of the presented method is affirmatively confirmed through comparison the result with that of conventional Mach-Zehnder based digital holographic method. Quantitative phase measurements on silica microspheres, onion skin and red blood cells ensure the validity of the method and its ability for monitoring nanometer-scale fluctuations of living cells, particularly in real-time.}


\section{Introduction}\label{sec1}

Optical imaging techniques are widely employed as a reliable and promising tool for biomedical research and diagnosis~\cite{Stephens:2003,Licha:2005,Sutton:2008,Robinson:2010,Majeed:2017,Sancho-Dura:2018}. One of the major limitations of conventional optical microscopes for identifying and investigating the transparent biological samples is their insufficient contrast, resulted from the low absorption and scattering of the transmitted light through them. Digital holographic (DH) microscope, as a well-established interferometric phase-contrast microscopy (IPM) tool, provides three dimensional information of objects through the numerical reconstruction of the recorded holograms~\cite{Marquet:2005,Rappaz:2005,Sheng:2006,Su:2013,Doblas:2016}. However, DH microscopy methods mostly rely on two-beam off-axis interference of light, induced by a beam interacting with the sample and a reference beam. In the dynamic measurements, different environmental effects between the beams, namely, the individual mechanical vibration of optical components, inevitably trigger some remnant time-dependent noises which reduce the quality of retrieved images. Recently, many efforts have been devoted toward developing new settings with the higher phase sensitivity such as self-reference or common-path interferometers (CPIs), through which the phase stability of the system is significantly enhanced as the two overlapping beams travel along nearly equivalent paths~\cite{Bon:2009,Singh:2012,Chhaniwal:2012,Shaked:2012,Girshovitz:2013,Bishitz:2014,Lee:2014,Calabuig:2014,Ebrahimi:2014,Mahajan:2015,Roitshtain:2016,Rawat:2017,Ebrahimi:2018,Sun:2018,Yaghoubi:2019}. Despite having the several advantages including simplicity and phase stability, many of the CPIs suffer from functional challenges which are imposed by loss of flux, inflexible off-axis angle, small field of view (FOV) limitation and precise optical alignment.

In this paper, we present a simple and portable quantitative phase microscope in which Fresnel diffraction pattern has been successfully used to precisely measure the optical path length information of transparent specimens such as living cells. The proposed method is high-contrast, very stable, and simple-to-align which make it feasible for integration with conventional microscopes at low-cost. The compactness of the method along with the capability of image acquisition in very short exposure times lead to the high measurement stability, which efficiently eliminates the effect of temporal phase noise on the reconstructed images. Moreover, the ability of modifying the fringe period independently from the FOV via rotating the plate, eliminates the necessity of phase-shifting procedure for investigating large FOVs.

Thus far, Fresnel diffraction from a phase step is studied~\cite{Amiri:2007,Tavassoly:2009,Salvdari:2017} and widely considered in the context of different metrological applications, e.g., the measurement of refractive indices of solids and liquids~\cite{Tavassoly:2010,Tavassoly:2012,Beygi:2015}, thickness of films and plates~\cite{Hassani:2016}, focal length of an imaging system~\cite{Dashtdar:2016}, etching rate of glass steps~\cite{Mahmoudi:2018}, and determination of the off-axis angle and the wavelength of light~\cite{Akhlaghi:2018}.

\section{Theory}\label{sec2}

When a coherent beam of light illuminates a plane-parallel phase plate, Fresnel diffraction pattern will appear near the boundary of the plate owing to an abrupt change in the phase of incident wave. The diffracted intensity in an arbitrary point $p$ in the observation plane can be obtained by the Fresnel-Kirchhoff integrals as~\cite{Amiri:2007,Tavassoly:2009}
\begin{align}    \label{e1}
I_{p}&=~I_{0}\,t\,\,\Big[\cos^{2}(\dfrac{\phi}{2})+2\,\sin^{2}(\dfrac{\phi}{2})\,\left(C^2(\omega)+S^2(\omega)\right)\\  \nonumber
&\mp \left(C(\omega)-S(\omega)\right) \sin\phi\Big]\\  \nonumber
& +\dfrac{I_{0}}{2}\,\Big[(1-t)^{2}\left(C^2(\omega)+S^2(\omega)+\dfrac{1}{2}\right)\\  \nonumber
&\pm (1-t^{2}) \left(C(\omega)+S(\omega)\right) \Big]
\end{align}
where $I_{0}$ is the illuminating intensity and $t$ is the amplitude transmission coefficient through the plate.
\begin{equation}\label{e2}
\phi =\dfrac{2 \pi}{\lambda}h \left[\sqrt{n^{2}-\sin ^{2}\theta}-\cos \theta\right]
\end{equation}
is the phase difference between the two beams of wavelength $\lambda$; one passing through the plate of refractive index $n$ and thickness $h$ at incident angle $\theta$ and the other propagating in air. The parameters $C({\omega})$ and $S(\omega)$ are the Fresnel Cosine and Sine integrals, where $\omega$ defines the upper limit of the integrals. The Fresnel integrals are constant along the lines parallel to the plate edge. According to Eq.~(\ref{e1}), the diffracted intensity is a periodic function of $\phi$ which could be varied by changing the angle of incidence.
\begin{figure}[!t]
\centering
\includegraphics[width= 0.7 \linewidth]{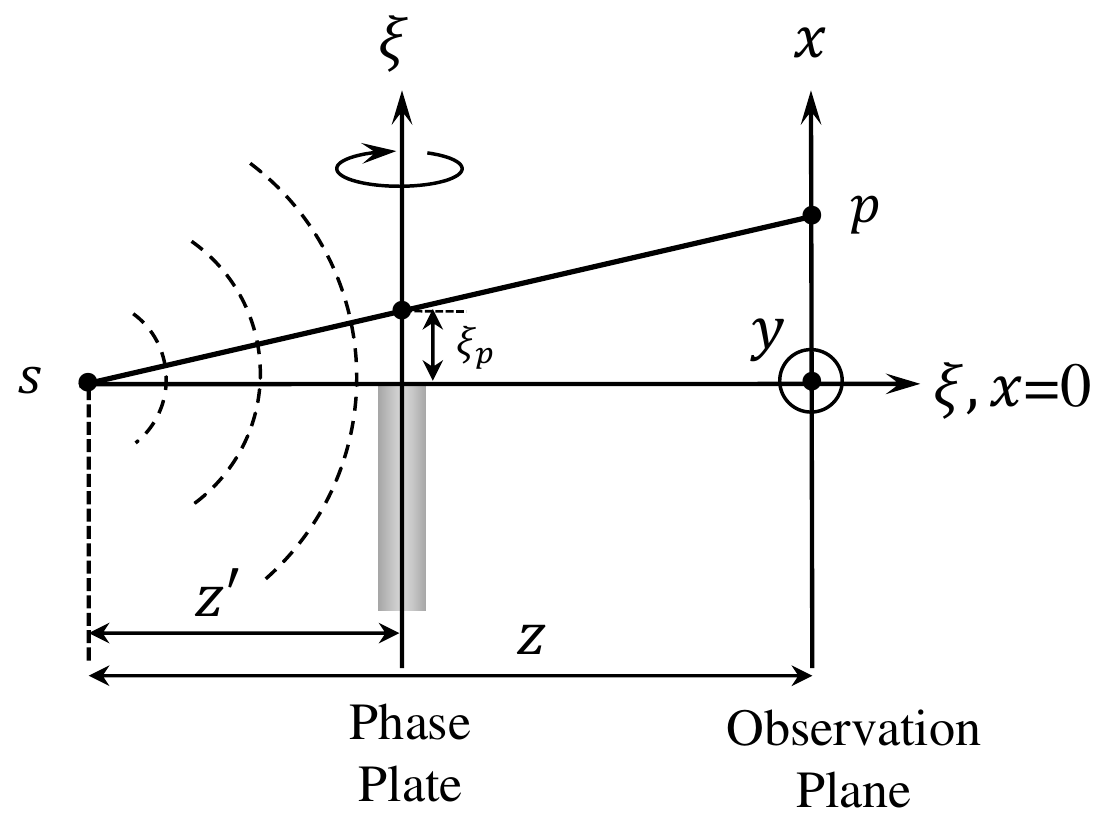}
\centering
zxc\caption{Side-view schematic of Fresnel diffraction of light from a phase plate.}
\label{f1}
\end{figure}
\begin{figure}[!b]
\centering
\includegraphics[width=\linewidth]{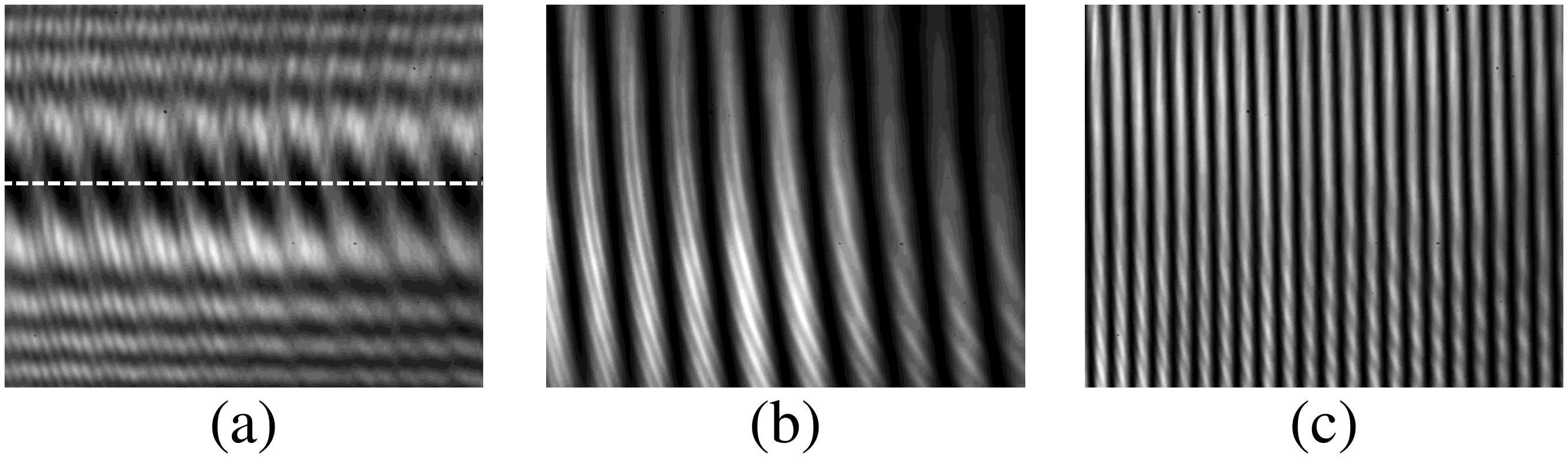}
\caption{Recorded diffraction pattern for maximum pixel resolution of CMOS, where the plate is: (a) a few centimeters far from $s$ point and is rotated by about $0.02$ radian, (b) a few millimeters far from $s$ point in a same rotation angle as (a), and (c) rotated by about $0.06$ radian in a same position situation as (b). The dashed line is related to the plate edge.}
\label{f2}
\end{figure}

In a spherical beam, the incident angle varies continuously along the plate edge that results in the periodic intensity pattern along the lines parallel to the edge. Figure~\ref{f1} illustrates the geometry of diffraction for a spherical wave, radiated by the point source $s$, from a phase plate. The plate and the observation plane ($x,y$) are located at distances $z^{\prime}$ and $z$ from the source, respectively. For an arbitrary point $p$ on the observation plane, $\omega$ is proportional to the location of the plate edge relative to the $s$ point, $\xi _{p}=x_{p} z^{\prime}/z$~\cite{Guenther:2015}. By placing the plate very close to the point source, i.e., $z^{\prime} \ll z$, the parameter $\xi _{p}$ (and consequently $\omega$) is negligible in comparison with $x_p$. That being the case, the intensity distribution in the directions parallel to the plate edge becomes akin to that of along the edge ,$x=0$, at which $C(0)=S(0)=0$. By this consideration, the intensity of diffraction pattern is given by
\begin{equation}\label{e3}
 I_{p}=\dfrac{I_{0}}{4}\left[ \left(1+t^{2} \right) +2t \cos \phi \right],
\end{equation}
The spatial frequency of these periodic fringes depends on the incident angle of the light as well as the height and refractive index of the step.
The diffracted intensity distribution of Eq.~(\ref{e3}) is similar to that of an interference phenomenon which could be numerically analyzed by Fourier transform method~\cite{Kreis:1986}.

In order to obtain the typical pattern of diffraction, a divergent beam of light, generated from an expanded beam of a He-Ne laser (Thorlabs, $632.8$ nm, $5$ mW) by a lens with a focal length of $100$ mm, illuminates the edge of the phase plate ($BK7$ slide, refractive index$=1.51$, thickness$=7.75$ mm). The diffraction pattern is recorded by a CMOS sensor (Basler acA1300-200um, $8$ bit dynamic range, $4.8$ $\mu$m pixel pitch) which is located at a distance of about 15 cm from the lens$\text{'}$ focal plane. Fig.~\ref{f2}(a) depicts the diffraction pattern at maximum resolution of the sensor ($1280\times 1024$ pixels), where the plate is rotated by about $0.02$ radian, for $z^{\prime}\approx z/3$.
By increasing the distance from the plate edge, high intensity variation as well as the contrast reduction of fringes are clearly obvious which are attributed to the oscillation of $C(\omega)$ and $S(\omega)$. These effects could be eliminated by locating the plate near to the focal point.
Figure~\ref{f2}(b) illustrates the diffraction pattern in the same FOV as Fig.~\ref{f2}(a) for $z^{\prime}$ of approximately a few millimeters.
By rotating the plate, the phase change and hence the fringe period varies without reduction of the FOV, as shown in Fig.~\ref{f2}(c) for about $0.06$ radian rotation angle. The few variations of contrast along $x$ direction, which is measured by $0.04$, demonstrates that $C(\omega)$ and $S(\omega)$ are approximately constant in the whole FOV recorded by the sensor. If a phase object is located in the path of the incoming light to the plate, the diffraction pattern varies depending on its height and refractive index. Thus, the phase information of object is achievable by measuring the fringe displacement.
\begin{figure}[!t]
\centering
\includegraphics[width= 0.7\linewidth]{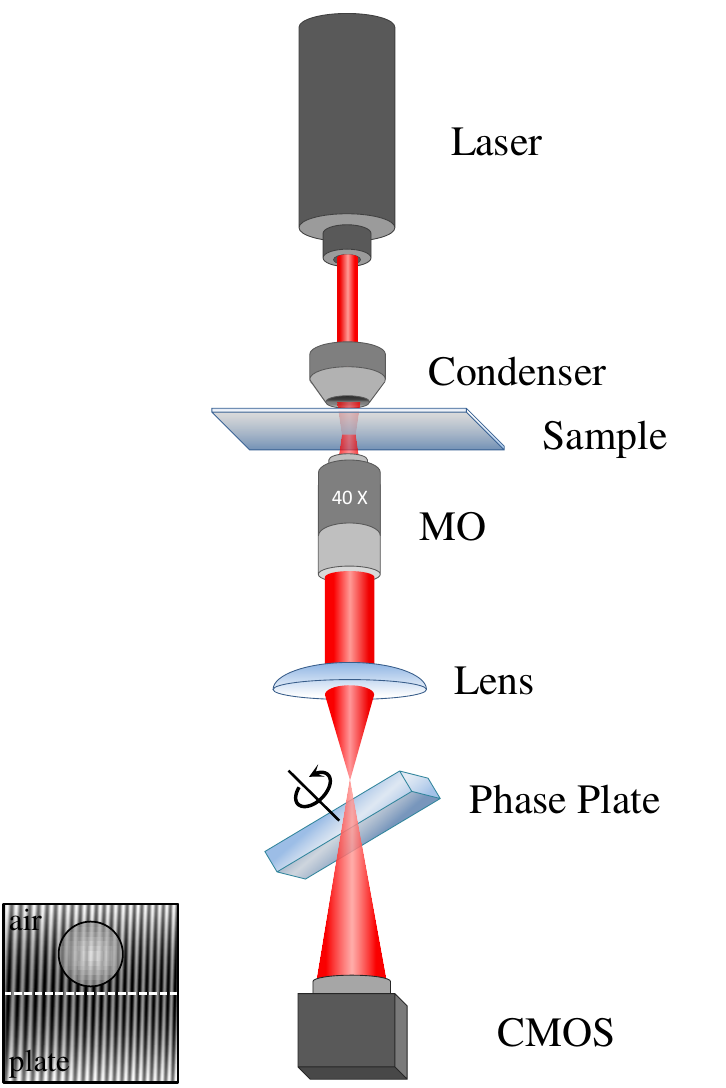}
\caption{Experimental setup of phase imaging system based on Fresnel diffraction. The left-bottom inset demonstrates the digital FOV, recorded by the sensor, in which the position of edge is shown by white-dashed line.}
\label{f3}
\end{figure}

\section{Experimental setup}\label{sec3}

The experimental diffraction-based imaging system is schematically shown in Fig.~\ref{f3}. The system comprises a conventional inverted microscope upgraded with the phase plate to attain a DH microscopy device. The He-Ne laser light is coupled to the imaging system, equipped with a condenser lens, microscope objective (MO) and the lens. The sample is imaged onto the sensor. The geometry of the plate relative to the focal plane and the sensor is similar to that in which Figs.~\ref{f2}(b) and \ref{f2}(c) are recorded. In the specimens which are not sparse enough, one could put the sample in half of the illumination spot relative to the plate edge to avoid overlapping of sample information and ghost image and this poses a half-FOV limitation.
\begin{figure}[!b]
\centering
\includegraphics[width= \linewidth]{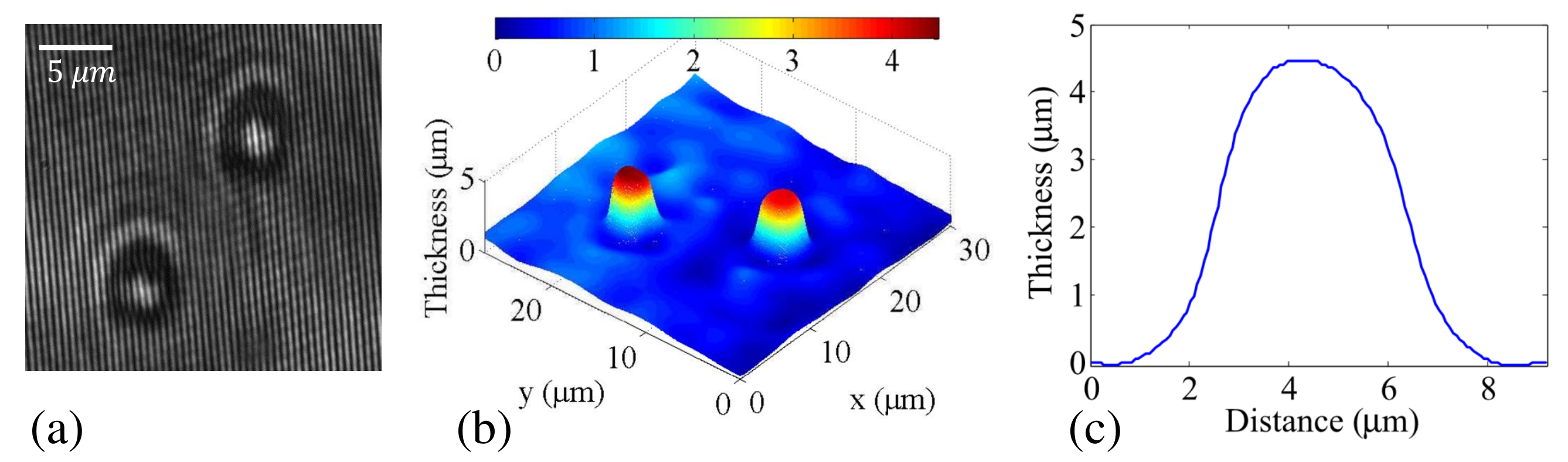}
\caption{(a) A hologram containing silica microspheres, obtained by the method. The inset demonstrates the Fourier domain. (b) 3D representation of thickness map. (c) Central cross-section of a single bead.}
\label{f4}
\end{figure}
\begin{figure}[!t]
\centering
\includegraphics[width= \linewidth]{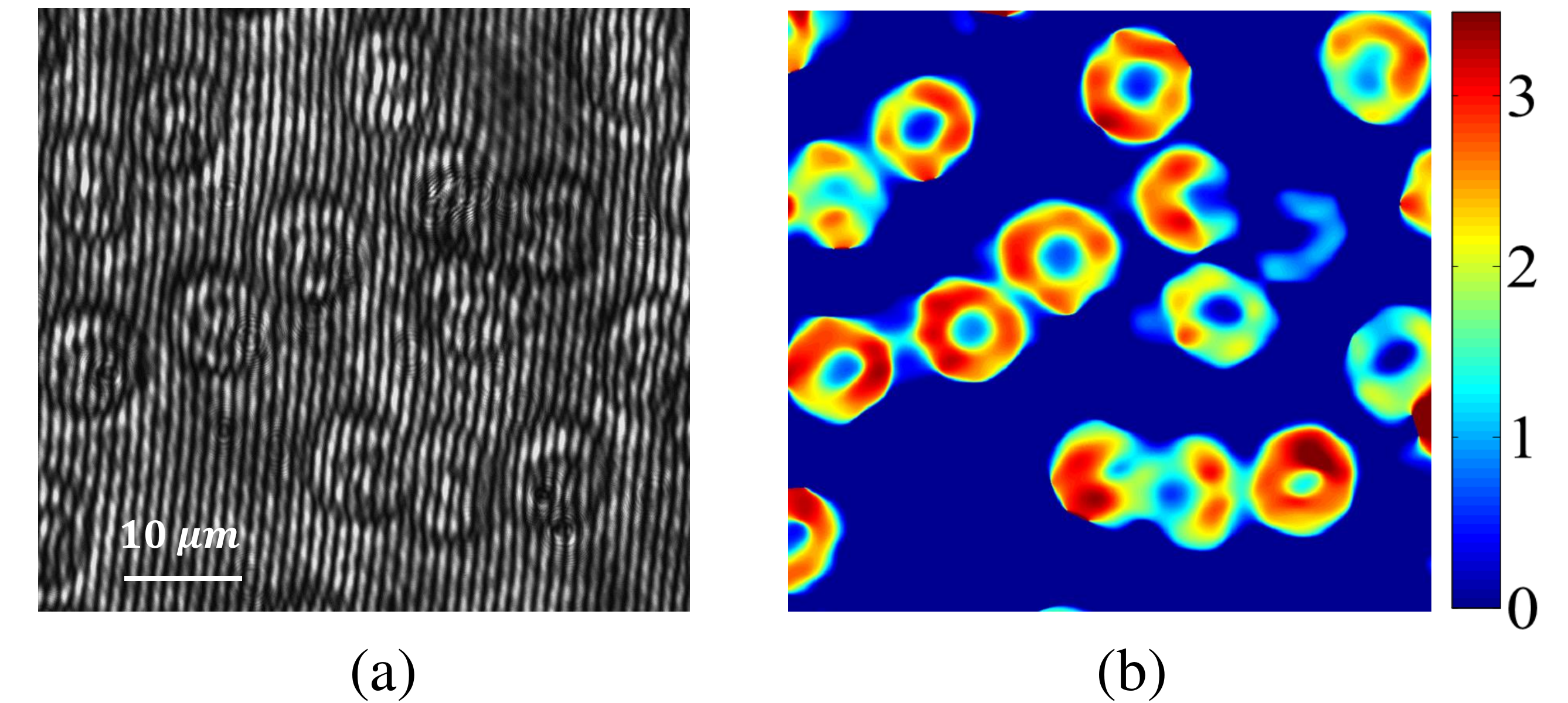}
\caption{(a) Hologram of RBCs. (b) The unwrapped phase distribution.}
\label{f5}
\end{figure}
\begin{figure}[!b]
\centering
\includegraphics[width= \linewidth]{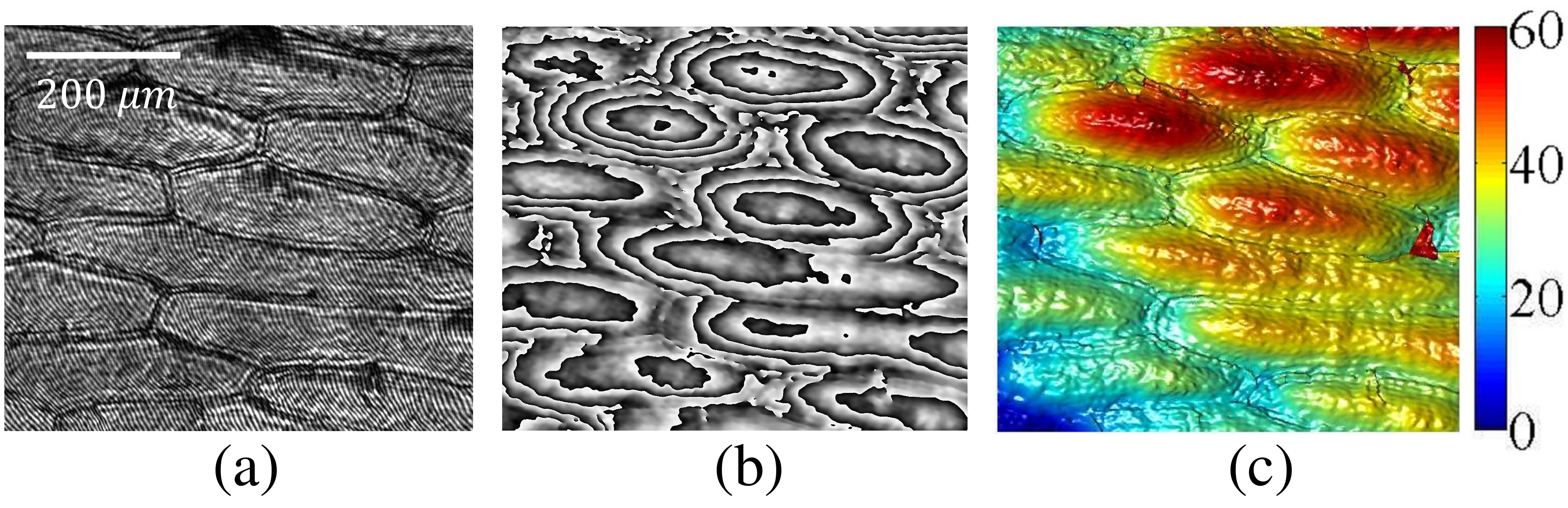}
\caption{(a) Hologram of onion skin cells. (b) Wrapped and (c) unwrapped reconstructed phase distributions.}
\label{f6}
\end{figure}

\section{Results and discussions}\label{sec4}

To experimentally characterize the capability of the proposed method, first silica microspheres (beads) with a diameter of $4.47$ $\mu$m, immersed in water are employed as the standard test samples. Figure~\ref{f4}(a) shows the recorded digital hologram of two beads using a $40\times$ (NA$=0.65$) microscope lens while the plate is rotated by about $0.19$ radian. The resulted fringe pattern is numerically reconstructed by Fourier transform method. By windowing one of the spectra in angular spectrum domain using a rectangular filter (see the inset of Fig.~\ref{f4}(a)) and applying an inverse Fourier transform of the filtered and shifted spectrum, the wrapped phase image is obtained. After subtracting a reference phase map from the object phase map, a Goldstein$\text{'}$s branch-cut unwrapping method is applied to reconstruct a continuous phase of object~\cite{Gutmann:2000}. The thickness distribution which is achieved from the net unwrapped phase information of silica beads, $d=\phi ^{\prime} \lambda /\left[2\pi (n_{O}-n_{m})\right]$, is illustrated in Fig.~\ref{f4}(b), where the refractive indices of the beads and the surrounding medium read as $n_{O}=1.42$ and $n_{m}=1.33$, respectively. The thickness cross-section across the diagonal of a single bead (Fig.~\ref{f4}(c)) implies that the resulted average microsphere size suggests a good agreement with the value reported by the manufacturer.

Next, to prove the potential applicability of the method for imaging biological samples which mostly have sparse populations, we used thin smear of human blood on a microscope slide. Figures~\ref{f5}(a) and \ref{f5}(b) show the recorded hologram and reconstructed phase distribution of red blood cells (RBCs), respectively. According to Fig.~\ref{f5}(b), as the two types of healthy, donut-shaped RBCs and disordered RBCs are clearly distinguishable, the proposed system may potentially be able of performing diagnosis of many types of diseases which cause the morphological modification of cells. As an additional demonstration of the configuration, confluent onion skin cells are imaged by a $10\times$ microscope lens, where half of the optical FOV is empty by laterally displacement of sample in relation to the edge of the phase plate. Figure~\ref{f6}(a) depicts the hologram of onion skin and Figs.~\ref{f6}(b) and \ref{f6}(c) show the wrapped and unwrapped phase profiles, respectively.
\begin{figure}[!t]
\centering
\includegraphics[width=\linewidth]{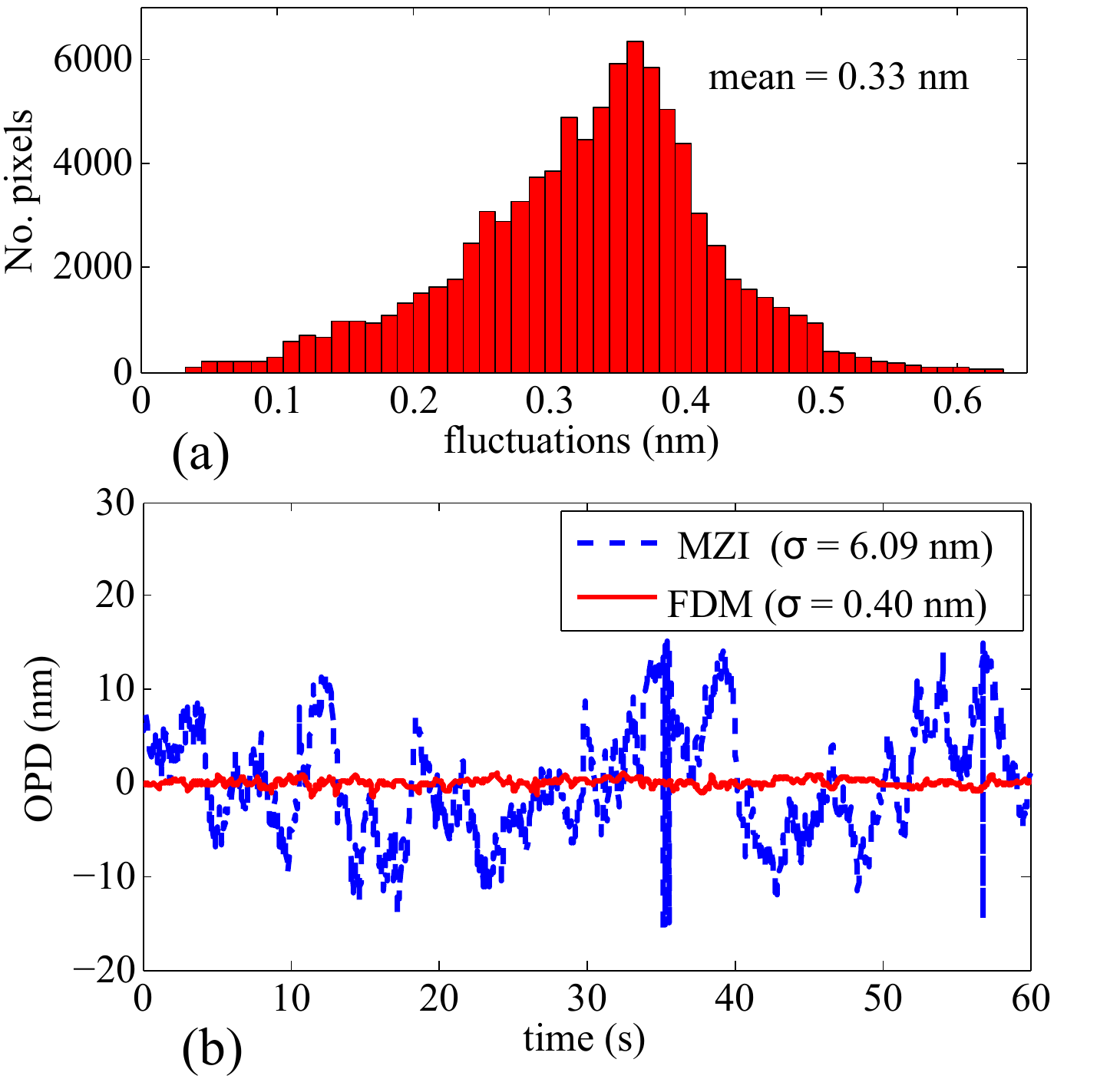}
\caption{(a) Histogram of standard deviations of the proposed setup fluctuations. (b) Temporal OPD of a diffraction-limited spot for both FDM and MZI method. $\sigma$ denotes the standard deviation of OPD.}
\label{f7}
\end{figure}
\begin{figure}[!t]
\centering
\includegraphics[width=0.6\linewidth]{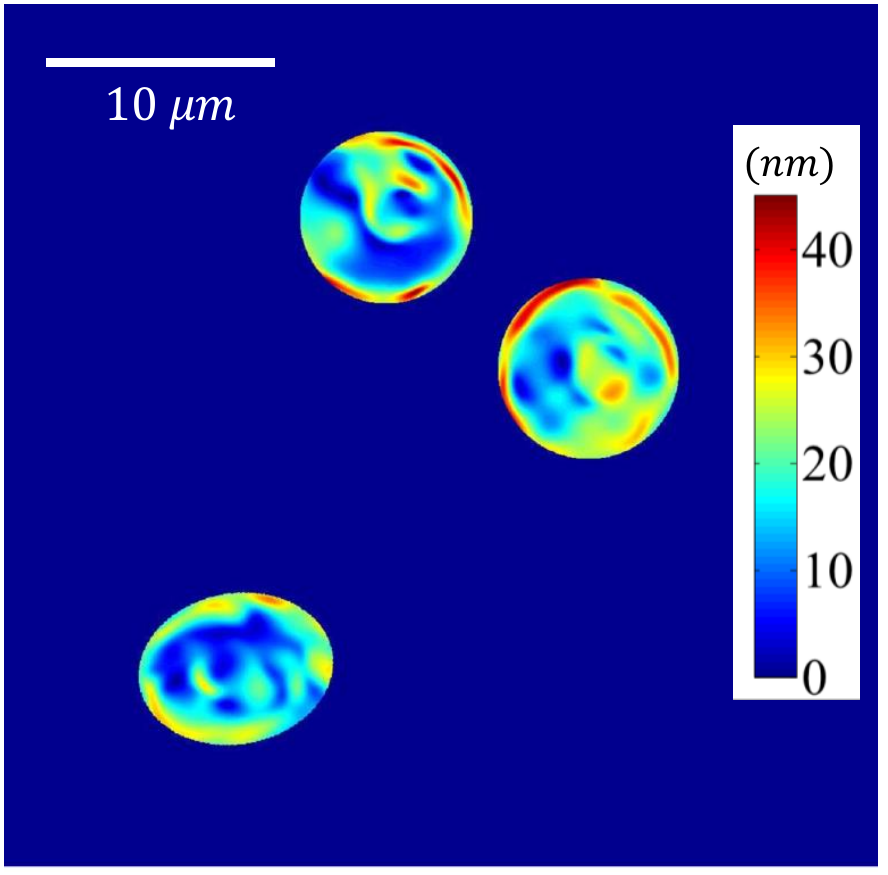}
\caption{The membrane fluctuation of RBCs obtained of $170$ temporally recorded holograms.}
\label{f8}
\end{figure}

To evaluate the temporal phase stability, as one of the most important features of a quantitative phase microscope, the number of $2460$ sample-free diffractograms are recorded at $41$ frames per second (fps) in the absence of vibration isolation. The fluctuation at each pixel is calculated through standard deviations of path length changes in the area of $300$ $\times$ $300$ pixels. The histogram of fluctuations, depicted in Fig.~\ref{f7}(a), reveals an approximately normal distribution with a mean value of $0.33$ nm. The mean value of the fluctuations could be considered as the temporal phase stability of the configuration. In order to compare the robustness of the Fresnel diffraction method (FDM) with a conventional IPM tool, the temporal optical path difference (OPD) of an approximately diffraction-limited spot containing $3$ $\times$ $3$ pixels is calculated for both a standard Mach-Zehnder interferometer (MZI) configuration and FDM (see Fig.~\ref{f7}(b)). The experiments are performed under identical conditions for a fair comparison. The OPD standard deviation of $0.40$ nm in the case of FDM proves more stable and noise-free configuration than MZI with OPD standard deviation of $6.09$ nm. Sub-nanometer temporal stability promotes the proposed method as a bright candidate for studying the dynamical phenomena such as nanometric cell membrane oscillations with high precision. In addition, because of removing the additional optical components such as mirrors and beam splitters, the proposed technique minimizes the loss of flux and allows for fast imaging with higher signal to noise ratio without necessity to a high-power laser which is an essential requirement in the fast and sensitive cell measurements. To further substantiate such arguments, we recorded holograms of RBCs suspended in distilled water and normal saline, positioned between a microscope slide and cover glass, at $170$ fps for $1$ s. The root mean square of dynamical thickness fluctuations for three individual cells multiplied by a binary mask is presented in Fig.~\ref{f8} that clearly displays the membrane fluctuation of the RBCs with a mean value of $23.95$ nm.

\section{Conclusions}\label{sec5}

In conclusion, we introduced Fresnel diffraction of the phase plate as a very robust, simple and cost-effective
optimistic tool for quantitative phase microscopy. In contrast to the conventional interferometry methods which take advantage of the light interference, here, diffraction pattern which is constructed by locating the edge of the plate in the optical path of the conventional microscope, is utilized for reconstructing the phase information of specimens. Easy adjustment of the fringe period decoupled from FOV, allows for single frame of acquisition. Comparing the measured stability result of Fresnel diffraction method (FDM) to that of a conventional Mach-Zehnder interferometer reveals that FDM has more than fifteen times better vibration resistance. In view of the low loss of flux, our setting paves the way to acquire superb-contrast fringes in the  short exposure times in order to detect fast dynamical phenomena. These features along with the ability of straightforward alignment ensure that FDM could easily be employed for a wide range of applications, specially in the case of living cells. The morphology of standard microspheres, RBCs and onion skin as a confluent specimen as well as the membrane fluctuation of RBCs are successfully investigated in order to validate the efficiency of the configuration.

%

\end{document}